\documentclass{epl}

\title{Promotion of cooperation induced by nonlinear attractive effect in spatial Prisoner's Dilemma game}
\author{Jian-Yue Guan\inst{1} \and Zhi-Xi Wu\inst{1} \and Zi-Gang Huang\inst{1} \and Xin-Jian Xu\inst{2} \and Ying-Hai Wang\inst{1}}
\institute
{\inst{1} Institute of Theoretical Physics, Lanzhou
University, Lanzhou Gansu 730000, China\\
  \inst{2} Department of Electronic Engineering, City University of
Hong Kong, Kowloon, Hong Kong, China }

\begin{document}

\maketitle

\begin{abstract}
We introduce nonlinear attractive effects into a spatial
Prisoner's Dilemma game where the players located on a square
lattice can either cooperate with their nearest neighbors or
defect. In every generation, each player updates its strategy by
firstly choosing one of the neighbors with a probability
proportional to $\mathcal{A}^\alpha$ denoting the attractiveness
of the neighbor, where $\mathcal{A}$ is the payoff collected by it
and $\alpha$ ($\geq$0) is a free parameter characterizing the
extent of the nonlinear effect; and then adopting its strategy
with a probability dependent on their payoff difference. Using
Monte Carlo simulations, we investigate the density $\rho_C$ of
cooperators in the stationary state for different values of
$\alpha$. It is shown that the introduction of such attractive
effect remarkably promotes the emergence and persistence of
cooperation over a wide range of the temptation to defect. In
particular, for large values of $\alpha$, i.e., strong nonlinear
attractive effects, the system exhibits two absorbing states (all
cooperators or all defectors) separated by an active state
(coexistence of cooperators and defectors) when varying the
temptation to defect. In the critical region where $\rho_C$ goes
to zero, the extinction behavior is power law-like $\rho_C$ $\sim$
$(b_c-b)^{\beta}$, where the exponent $\beta$ accords
approximatively with the critical exponent ($\beta\approx0.584$)
of the two-dimensional directed percolation and depends weakly on
the value of $\alpha$.

\end{abstract}

\section{Introduction}

Cooperation plays an important role in real world, ranging from
biological systems to economic and social systems
\cite{cooperation}. Scientists from many different fields of
natural and social sciences often resort to Evolutionary Game
Theory \cite{evolution,game} as a common mathematical framework
and the prisoner's dilemma game (PDG) as a metaphor for studying
cooperation between unrelated individuals \cite{game}. The
original PDG describes the pairwise interactions of individuals
with two behavioral options: the two players must simultaneously
decide whether to cooperate or to defect. For mutual cooperation
both players receive the rewards $R$, but only the punishment $P$
for mutual defection. A defector exploiting a cooperator gets an
amount $T$ (temptation to defect) and the exploited cooperator
receives $S$ (sucker's payoff). These elements satisfy the
following two conditions: $T>R>P>S$ and $2R>T+S$. It is easy to
see that defection is the better choice irrespective of the
opponent's decision. Thus, the undesired outcome of mutual
defection emerges in well-mixed populations \cite{population},
which has inspired numerous investigations of suitable extensions
that enable cooperative behavior to emerge and persist.

Some previous works have suggested several mechanisms (e.g., kin
selection \cite{jtb}, the introduction of \lq\lq tit-for-tat\rq\rq
\cite{basic,nature1} strategy, and voluntary participation
\cite{science,szabo1,szabo2}) to facilitate the emergence and
persistence of cooperation in the populations. The spatial
versions \cite{ijbc1,ijbc2} of the evolutionary PDGs can explain
the maintenance of cooperation for the iterated games with a
limited range of interaction if the players follow one of the two
simplest strategies (defection ($D$) and cooperation ($C$)).
Recently, the effect of heterogeneous influence of different
individuals on the maintenance of cooperative behavior has been
studied on regular small-world networks \cite{cpl}. Ren \emph{et
al.} have studied the evolutionary PDG and the snowdrift game with
preferential learning mechanism on the Barab\'{a}si-Albert
networks \cite{physics}. In the evolutionary games the players
wish to maximize their total payoffs, coming from PDGs with the
neighbors, by adopting either the deterministic rule introduced by
Nowak and May \cite{ijbc1,nature2} or the stochastic evolutionary
rule by Szab\'{o} and T\H{o}ke \cite{szabo3}.

In the present work, we make further studies of the evolutionary
PDG on square lattice mainly according to the stochastic update
rule. It is natural to consider that different individuals may
have different attractiveness in social systems, so when updating
their strategies, the individuals may not completely randomly
choose a neighbor to refer to. Here, we introduce the nonlinear
attractive effect into the game (see the model below).
Interestingly, we find that the introduction of this effect can
remarkably promote cooperative behavior in the PDG in comparison
with the random choice case on square lattice.

\begin{figure}
\scalebox{0.62}[0.5]{\includegraphics{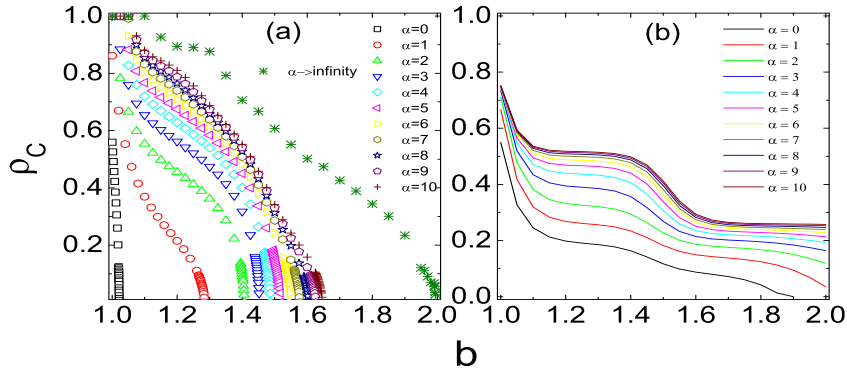}}
\caption{\label{fig:epsart} (color online). Monte Carlo results
(a) and theoretical analysis (b) for cooperator density $\rho_C$
in the steady state as a function of the temptation to defect $b$
for several different values of $\alpha$ (see the plot). The
pair-approximation correctly predicts the trends that $\rho_C$
changes with $b$ and $\alpha$, but significantly overestimates the
extinction thresholds in contrast to the simulation results (see
the text for details).}\label{fig1}
\end{figure}

\begin{figure}
\includegraphics[width=12.5cm]{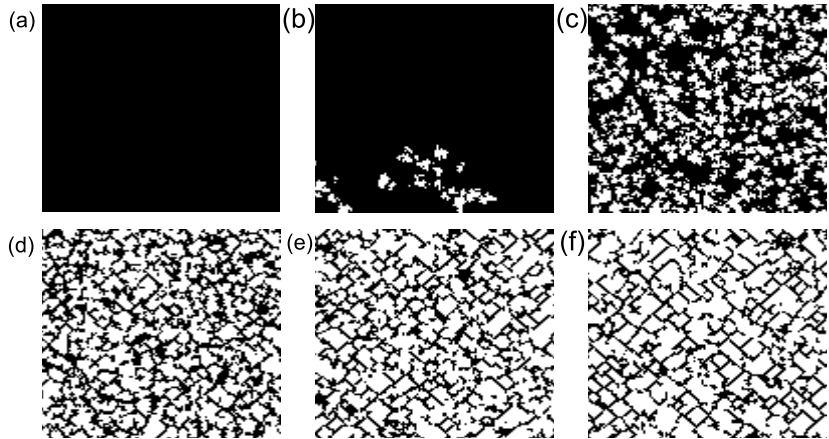}

\caption{A series of snapshots of typical distributions of
cooperators (white) and defectors (black) on square lattice for
$b=1.283$ (the value just below the extinction threshold
$b_{c1}\approx1.284$ when $\alpha=1$) for several different values
of $\alpha$: (a) $\alpha=0$, (b) $\alpha=1$, (c) $\alpha=2$, (d)
$\alpha=5$, (e) $\alpha=8$, (f) $\alpha=10$.}\label{fig2}
\end{figure}

\begin{figure}
\includegraphics[width=12.5cm]{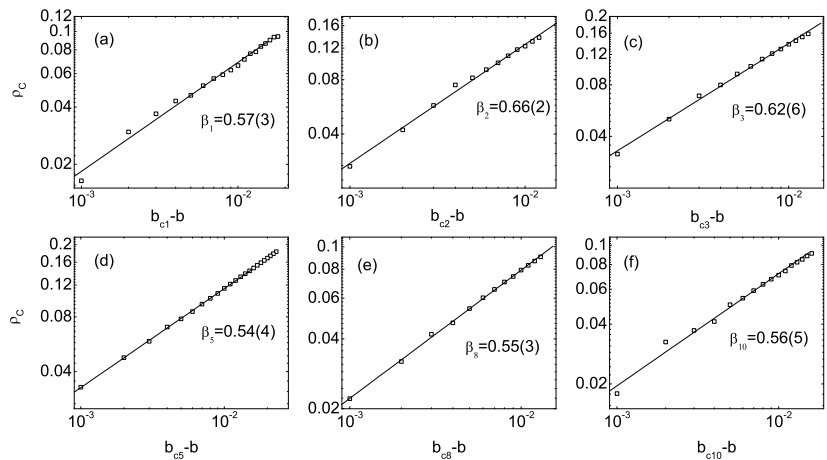}

\caption{Log-log plots of the average cooperator density $\rho_C$
as a function of the distance to the extinction threshold
$b_{c\alpha}-b$ for several values of $\alpha$: (a) $\alpha=1$,
$b_{c1}\approx1.284$, (b) $\alpha=2$, $b_{c2}\approx1.407$, (c)
$\alpha=3$, $b_{c3}\approx1.453$, (d) $\alpha=5$,
$b_{c5}\approx1.518$, (e) $\alpha=8$, $b_{c8}\approx1.603$, (f)
$\alpha=10$, $b_{c10}\approx1.646$. The solid lines, the power
laws $\sim(b_{c\alpha}-b)^{\beta_{\alpha}}$ fit the data
correspondence with the exponents of $\beta_{\alpha}$ (the
detailed values are given in the plots, where the figures between
parentheses indicate the statistical uncertainties of the last
digit).}\label{fig3}
\end{figure}

\begin{figure}
\scalebox{0.80}[0.7]{\includegraphics{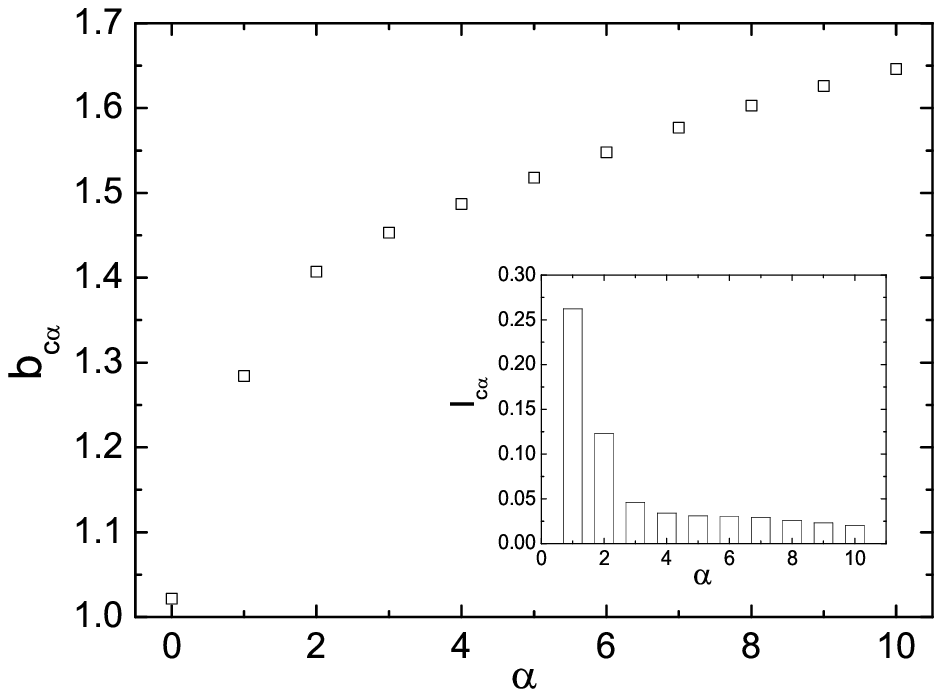}}
\caption{\label{fig:epsart}. The extinction threshold
$b_{c\alpha}$ changes with $\alpha$ (from zero to ten). Inset
shows that the increment $I_{c\alpha}$ of the extinction threshold
$b_{c\alpha}$ changes with $\alpha$ (see the text).}\label{fig4}
\end{figure}

\section{The model}

We consider the evolutionary PDG on square lattice with periodic
boundary conditions. Each player interacts only with its four
nearest neighbors (self-interaction is excluded) \cite{ajp}, and
collects payoffs dependent on the payoff-matrix parameters. The
total payoff of a certain player is the sum over all its
interactions. We have inspected that if every player interacts
with its first and second nearest neighbors or the
self-interactions are included, the qualitative results are
unchanged. Following common practices \cite{nature2,szabo3}, we
start by rescaling the game to make $T=b$, $R=1$, and $P=S=0$,
where $b$ represents the advantage of defectors over cooperators
\cite{nature2}, being typically constrained to the interval
$1.0<b<2.0$, such that it depends on a single parameter $b$. We
have checked that the qualitative results do not change if we make
$S=-\epsilon<0 (\epsilon\ll1)$ in order to strictly enforce a PD
setting.

During the evolutionary process, each player is allowed to select
one of its neighbors as a reference with a probability
proportional to the neighbors' attractiveness, and then decides
whether to change its strategy or not dependent on their payoff
difference. We define the selection probability $P_{x\rightarrow
y}$ of $x$ selecting a neighbor $y$ as
\begin{equation}
P_{x\rightarrow y}=\frac{\mathcal{A}_y^\alpha}{\sum_{z\in\Omega_x}
\mathcal{A}_z^\alpha}, \label{rule1}
\end{equation}
where the numerator denotes the attractiveness of the neighbor
$y$, and $\alpha$ is a tunable parameter describing the extent of
the nonlinear effect, and $\mathcal{A}_y$ is the total payoff of
that neighbor. The denominator is the sum of attractiveness that
runs over all neighbors of $x$. The basic ingredient which
determines the choice of one neighbor is the \emph{selection
kernel} $\mathcal{A}^\alpha$. On general grounds, this
\emph{selection kernel} should be a nondecreasing function of
$\mathcal{A}$, namely individuals with better performance may have
much stronger attractiveness than the average individual. We note
that the selection probability depends only on the extent of a
nonlinear effect $\alpha$ since the total payoff collected by any
player satisfies $\mathcal{A}\geq0$ in the present model. Thus we
will consider the model with $\alpha \in [0,\infty)$. For
$\alpha=0$, the neighbor is randomly selected so that the game is
reduced to the original one in Refs. \cite{szabo3,ajp}. The case
$\alpha=1$ leads to the proportional selection rule (exclude the
player itself) \cite{preph}. While in the limit of
$\alpha\rightarrow\infty$, the neighbor whose payoff is the
highest among the neighbors is selected, which resembles to the
deterministic selection rule \cite{ijbc1,nature2}. For other
values of $\alpha$, the attractiveness of the neighbors is a
nonlinear function of their total payoffs. In this way, we
consider the general situations of the nonlinear attractive effect
on the dynamical behavior of the game.

The player $x$ adopts the selected $y$ neighbor's strategy in the
next round with a probability depending on their total payoff
difference presented in Ref. \cite{szabo3,ajp,szabo4,szabo5} as
\begin{equation}
W(x\leftarrow
y)=\frac{1}{1+exp[(\mathcal{A}_x-\mathcal{A}_y)/\kappa]},
\label{rule2}
\end{equation}
where $\mathcal{A}_x$, $\mathcal{A}_y$ denote the total payoffs of
individuals $x$ and $y$ respectively, and $\kappa$ characterizes
the noise effects, including fluctuations in payoffs, errors in
decision, individual trials, etc. The effect of noise has been
reported by Szab\'{o} \emph{et al.} \cite{szabo5}. In this paper,
we make $\kappa=0.1$. Qualitatively, the results remain unaffected
when changing the parameter $\kappa$.
\section{Simulations and analysis}

Simulations were carried out for a population of $N=400\times400$
individuals. We study the key quantity of cooperator density
$\rho_C$ in the steady state. Initially, the two strategies of $C$
and $D$ are randomly distributed among the individuals with equal
probability $1/2$. The above model was simulated with synchronous
updating \cite{nature3}. No qualitative changes occur if we adopt
an asynchronous updating \cite{szabo3}. Eventually, the system
reaches a dynamic equilibrium state. The simulation results were
obtained by averaging over the last $5000$ Monte Carlo time steps
of the total $50000$.

Fig. \ref{fig1} shows the results of both simulations and
theoretical analysis of $\rho_C$ when increasing $b$ for several
different values of $\alpha$ (see the plot). We can find that,
compared with the well-mixed situation, if $b$ is sufficiently
small, cooperators can persist in spatial settings in the case of
$\alpha=0$, which indicates that spatial structure can promote
cooperation \cite{nature2}. For $b>b_{c0}$ ($\approx1.0217$),
where $b_{c0}$ is the extinction threshold of cooperators when
$\alpha=0$, the benefits of spatial clustering are no longer
sufficient to offset the losses along the boundary, hence the
cooperators vanish \cite{ajp}. For each positive value of
$\alpha$, the system evolves to the absorbing state of all
defectors at certain values of $b$. The extinction threshold of
cooperators $b_{c\alpha}$ clearly increases with $\alpha$, which
indicates that the emergence of cooperation is enhanced.

We know that cooperators survive by forming compact clusters and
thus cooperators along the boundary can outweigh their losses
against defectors by gains from interactions within the cluster
\cite{ajp}. The payoffs collected by the inner cooperators are, in
most cases, larger than the boundary defectors. When considering
the attractiveness of the individuals, near the extinction
threshold, for cooperator-clusters which are surrounded by
defectors, the cooperators along the boundary can keep their
cooperative states more easily under stronger preferential
selection effect according to the dynamic updating Eq.
(\ref{rule1}) and Eq. (\ref{rule2}). Thus the strong
attractiveness of individuals can favor the spreading of
cooperators, hence promote the persistence of cooperation. We can
also see that, for very large $\alpha$, the homogeneous
cooperation state ($\rho_C=1$) emerges when $b$ is very small.
Since both the small temptation to defect and the strong nonlinear
attractive effect are advantageous to the persistence of
cooperation, it is not surprising that $\rho_C$ approaches the
maximal fraction $1$. Moreover, for the same value of $b$,
$\rho_C$ obviously increases with $\alpha$. The existence of
strong nonlinear effect can facilitate the formation of cooperator
clusters, hence enhance the persistence of cooperators (see Fig.
\ref{fig2}). In addition, the cooperator density $\rho_C$ and the
extinction thresholds $b_{c\alpha}$ change more slowly with
increasing $\alpha$. We will consider this point in the following.
The pair approximation method, which models the frequency of
strategy pairs rather than that of strategies, is usually regarded
as an analytical approximation of the spatial dynamics \cite{ajp}.
It is worth noting that the results obtained by pair approximation
are directly associated with the local topological structure of
the players and the strategy updating dynamics. Whether the
strategy updating is implemented synchronously or asynchronously
is inessential \cite{nature3}. Therefore we can apply the pair
approximation method to predict qualitatively the evolving
behavior of $\rho_C$. Here we modify the original method by
introducing the turnable parameter $\alpha$, i.e., we substitute
the new transition probability $f'(P_{B}-P_{A})$ for the original
transition probability $f(P_{B}-P_{A})$ in Eq. (A1) in ref.[18],
where $f'(P_{B}-P_{A})=f(P_{B}-P_{A})\frac{P_{B}^{\alpha}}{\sum
P_{B}^{\alpha}}$ and the denominator of the fraction denotes the
sum of all possible values of $P_{B}^{\alpha}$ when $\alpha$ is
certain. Then we adjust the Eq. (A2a) and Eq. (A2b) in
ref.\cite{ajp} by substituting $f'$ for $f$. The equilibrium
values are obtained by numerical integration. We can see from Fig.
\ref{fig1} that the pair approximation correctly predicts the
trends, that is, the changes of cooperation for $b$ and $\alpha$.
However, it is unable to estimate exactly the extinction
thresholds of cooperator density, namely, it overestimates the
extinction thresholds (see the plot).

A series of snapshots for several different values of $\alpha$ are
shown in Fig. \ref{fig2} for the same value of $b$.These snapshots
are a $120\times120$ portion of the full $400\times400$ lattice.
We can find that, for random selection case $\alpha=0$,
cooperators are doomed and defectors reign because the value
$b=1.283$ is larger than $b_{c0}\approx1.0217$ (Fig.
\ref{fig2}(a)). But for $\alpha=1$, when $b$ is just below
$b_{c1}$, the cooperators can survive by forming compact clusters
which minimize the exploitation by defectors (see Fig.
\ref{fig2}(b)). For larger $\alpha$, i.e., stronger nonlinear
attractive effect, more cooperator clusters emerge, which
illustrates that cooperators can survive against the invading of
defectors more easily. On the contrary, the defector clusters
decrease with the value of $\alpha$. Interestingly, the spatial
patterns adopted by defectors are completely different from that
of cooperators when they are the minority in the populations. As
shown in Fig. \ref{fig2} (f), defectors exist in the fashion of
zigzag pattern (or step-like). Because if defector clusters are
surrounded by cooperators, in the subsequent generations, the
defectors along the boundary would transform probably to
cooperators according to the selection rule with nonlinear
attractive effects Eq. (\ref{rule1}) and updating rule Eq.
(\ref{rule2}). Eventually, defectors exist in zigzag pattern from
which they can benefit maximumly when interacting with their
cooperator neighbors (even can do better than a cooperator
surrounded by four cooperators, so that when updating, they \lq\lq
always\rq\rq ask their defector neighbors for strategy
transformation), which in return makes the zigzag pattern stably
in the evolution of the game.

We also investigate the divergence behaviors of cooperators near
the extinction thresholds $b_{c\alpha}$ for several values of
$\alpha$ (see Fig. \ref{fig3}). We constrained the system size as
$800\times800$ individuals and the results were obtained by
averaging over the last $10000$ time steps of the total $150000$.
According to our simulations, near $b_{c\alpha}$ the average
fraction of cooperators vanishes as power-law like behavior
$\rho_c\sim(b_{c\alpha}-b)^{\beta_{\alpha}}$, where
$\beta_{\alpha}$ are a set of exponents corresponding with the
value of $\alpha$ (see the plots for detailed values). In physics,
such thresholds are usually associated with phase transitions, and
indeed, the transitions from persistent levels of cooperation
($b<b_{c\alpha}$) to absorbing states of defection
($b>b_{c\alpha}$) bear the hallmarks of critical phase
transitions. These values of $\beta_{\alpha}$ are nearly
consistent with the critical exponent ($\beta\approx0.584$) of the
two-dimensional directed percolation \cite{Hinrichsen} and depend
weakly on $\alpha$. The estimated errors of $\beta_{1}$,
$\beta_{2}$, $\beta_{3}$, $\beta_{5}$, $\beta_{8}$, $\beta_{10}$
are $0.019, 0.133, 0.072, 0.068, 0.053, 0.033$ respectively. We
consider that the large errors are due to the limits of
computational conditions.

The variation of extinction threshold $b_{c\alpha}$ with the value
of $\alpha$ (from zero to ten) is shown in Fig. \ref{fig4}. We can
see that the extinction threshold clearly increases with the value
of $\alpha$. Now we define a quantity characterizing the increment
of the extinction threshold as
$I_{c\alpha}=b_{c\alpha}-b_{c(\alpha-1)}$. We note that this
quantity decreases with the value of $\alpha$ (inset in Fig.
\ref{fig4}). This indicates that, as $\alpha\rightarrow\infty$,
the increment $I_{c\alpha}$ will approach to the minimal value
$0$. In other words, the extinction threshold $b_{c\alpha}$ will
tend towards the maximal value $b_{c\infty}$, where
$b_{c\infty}\approx1.995$ is the extinction threshold when
$\alpha\rightarrow\infty$, i.e., when the neighbor whose payoff is
the highest of the neighboring is selected to refer to.

\section{Conclusions}

In summary, we have investigated the promotion of cooperation in
the context of evolutionary PDG resulting from the nonlinear
attractive effect of the neighbors on square lattice. A nonlinear
function $\mathcal{A}^{\alpha}$, in terms of the performance of
the players, is used as an estimator of their attractiveness. We
have considered the general situations and shown that, compared
with the random selection case, the introduction of the nonlinear
attractive effect can remarkably promote the cooperative behavior
over a wide range of $b$. Particularly, the stronger the extent of
the nonlinear effect is, the more prominent the cooperative
behavior will be, and for some large $\alpha$ values, a
homogeneous state of all cooperators can emerge. Interestingly,
the spatial patterns adopted by cooperators and defectors are
completely different when they are the minority in the
populations: Cooperators can survive by forming compact clusters,
and along the boundary, cooperators can outweigh their losses
against defectors by gains from interactions within the cluster;
Whereas defectors exist in the way of zigzag pattern (or
step-like), from which defectors can benefit maximumly when
interacting with their cooperator neighbors. The extinction of
cooperators under harsh conditions when $b \rightarrow
b_{c\alpha}$ displays a power law-like behavior
$\rho_C\sim(b_{c\alpha}-b)^{\beta}$. The introduction of the
nonlinear attractive effect can partially resolve the dilemma of
cooperation and may shed new lights on the evolution of
cooperation in the society.

\acknowledgments This work was supported by the Fundamental
Research Fund for Physics and Mathematics of Lanzhou University
under Grant No. Lzu05008.

\end{document}